\documentclass[12pt,aip]{article}
\usepackage{graphicx,epsfig,amsmath,xcolor,ulem,nccmath,lipsum,tikz,array,lineno}

\usepackage[colorlinks = true,
            linkcolor = blue,
            urlcolor  = blue,
            citecolor = blue,
            anchorcolor = blue]{hyperref}

\usepackage[nodisplayskipstretch]{setspace}
\begin{document}

\begin{center}
{\bf USING MACHINE LEARNING TO ASSESS SHORT TERM \\
\ \\
CAUSAL DEPENDENCE AND INFER NETWORK LINKS} \\
\ \\
Amitava Banerjee$^1$, Jaideep Pathak$^1$, Rajarshi Roy$^{1,2}$,
Juan G. Restrepo$^3$, and 

Edward Ott$^{1,4}$
\end{center}
1. Department of Physics and Institute for Research in Electronics and Applied Physics, University of Maryland, College Park, Maryland 20742, U.S.A.\\
2. Institute for Physical Science and Technology, University of Maryland, College Park, Maryland 20742, U.S.A. \\
3.  Department of Applied Mathematics, University of Colorado, Boulder, Colorado 80309, U.S.A. \\
4.  Department of Electrical and Computer Engineering, University
of Maryland, College Park, Maryland 20742, U.S.A.

{\bf Date:} December 2, 2019

\renewcommand{\baselinestretch}{2}
\small\normalsize

%\linenumbers

{\bf
\underline{Abstract}} \\
We introduce and test a general machine-learning-based technique for the inference of short term causal dependence between state variables of an unknown dynamical system from time series measurements of its state variables. Our technique leverages the results of a machine learning process for short time prediction to achieve our goal. The basic idea is to use the machine learning to estimate the elements of the Jacobian matrix of the dynamical flow along an orbit. The type of machine learning that we employ is reservoir computing. We present numerical tests on link inference of a network of interacting dynamical nodes. It is seen that dynamical noise can greatly enhance the effectiveness of our technique, while observational noise degrades the  effectiveness. We believe that the competition between these two opposing types of noise will be the key factor determining the success of causal inference in many of the most important application situations.

\newpage

{\bf
 
The general problem of determining causal dependencies in an
unknown time evolving system from time series observations is of great
interest in many fields.  Examples include inferring neuronal
connections from spiking data, deducing causal dependencies between
genes from expression data, discovering long spatial range
influences in climate variations, etc.  Previous work has often tackled
such problems by consideration of correlations, prediction impact,
or information transfer metrics. Here we propose a new method that leverages the potential ability of machine learning to perform predictive and interpretive tasks and uses this to extract information on causal dependence. We test our method on model complex systems consisting of networks of many interconnected dynamical units. These tests show that machine learning offers a unique and potentially highly effective approach to the general problem of causal inference.}

\noindent {\bf I. INTRODUCTION}

The core goal of science is often described to be generalization
from observations to \\ understanding,\textsuperscript{\cite{r1}}  commonly embodied in
predictive theories.  Related to this is the
desire to use measured data to infer necessary properties and
structure of any description consistent with
a given class of observations.  On the other hand, it has recently
emerged that machine learning (ML) is capable of effectively
performing a wide range of interpretive and
predictive tasks on data.\textsuperscript{\cite{r2}}  Thus it is natural
to ask whether machine learning might be useful for the common
scientific goal of discovering structural properties of a system
from data generated by that system.  In this paper we consider an
important, widely applicable class of such tasks.  Specifically,
we consider the use of machine learning to address two goals. 

{\it Goal
(i): Determine whether or not a state variable of a time evolving system causally
influences another state variable.
\\ Goal (ii): Determine  the `strength' of such causal influences.} 

In the terminology of ML, Goal (i) is referred to as ``classification ML,'' and Goal (ii) is referred to as ``regression ML.''
These goals have previously been of great
interest in many applications (e.g., economics,\textsuperscript{\cite{r3}}
neuroscience,\textsuperscript{\cite{r4}} genomics,\textsuperscript{\cite{r4b}} climate,\textsuperscript{\cite{r5a}} etc.). Many past approaches have, for example, been based upon the concepts of
 prediction impact,\textsuperscript{\cite{r3, r4}} correlation,\textsuperscript{\cite{r5,r5new,r5new2}} information
transfer,\textsuperscript{\cite{r6,r7a}} and direct physical perturbations\textsuperscript{\cite{r7c,r7new}}. Other previous works have investigated the inference of network links from time series of node states assuming some prior knowledge of the form of the network system and using that knowledge in a fitting procedure to determine links\textsuperscript{\cite{r5new2,r7,r7b,r7g,rnew4}}. In addition, some recent papers address network link inference from data via techniques based on delay coordinate embedding,\textsuperscript{\cite{r7b}} random forest methods,\textsuperscript{\cite{r7d}} network embedding algorithms\textsuperscript{\cite{r7e}} and feature ranking\textsuperscript{\cite{r7f}}. In this paper, we introduce a technique that makes the use of an ML training
process in performing predictive and interpretive tasks and attempts to use it to extract information about causal dependences. In particular, here we use a particular type of machine learning (ML) called reservoir computing, an efficient method of time series analysis which has previously been successfully used for different tasks, e.g., prediction of chaotic dynamics\textsuperscript{\cite{r12,r13,r14}} and speech recognition\textsuperscript{\cite{r21,r22}} to mention a few. In our case, a ``reservoir'' dynamical system is trained such that it becomes synchronized to a training time series data set from the unknown system of interest. The trained reservoir system is then able to provide an estimation of the response to perturbations in different parts of the original system, thus yielding information about causal dependencies in the actual system.  We will show that this ML-based technique offers a unique and
potentially highly effective approach to determining causal dependences. Furthermore, the presence of dynamical noise (either naturally present or intentionally injected) can very greatly improve the ability to infer causality,\textsuperscript{\cite{r7,r7b}} while, in contrast, observational noise degrades inference.

\newpage
\noindent {\bf II. SHORT TERM CAUSAL DEPENDENCE (STCD)}

We begin by considering the very general case of an evolving, deterministic, dynamical system whose
state at time $t$ is represented by the $M$-dimensional vector
${\bf z}(t) = [z_1(t), z_2(t), \ldots, z_M(t)]^T$, where ${\bf z}(t)$ evolves via a system
of $M$ differential equations, $d{\bf z}(t)/dt = {\bf
F}({\bf z}(t))$, and has
reached a statistically steady dynamical state (perhaps chaotic).
In this context, we frame the issue of causality as follows:
Will a perturbation at time $t$ applied to a component $z_i$ of
the state vector ${\bf z}(t)$ (i.e., $z_i(t) \rightarrow z_i(t) +
\delta z_i(t)$) lead to a subsequent change at a slightly later time, $t + \tau$, of another scalar component
$z_j$ (i.e., $z_j(t + \tau) \rightarrow z_j(t + \tau) + \delta z_j
(t + \tau)$); and how can we quantify the strength of this
dependence? This formulation might suggest comparison of the
evolutions of ${\bf z}(t)$ that result from two identical systems,
one with, and the other without, application of the perturbation.
However, we will be interested in the typical situation in which
such a comparison is not possible, and one can only passively
observe (measure) the state ${\bf z}(t)$ of the (single) system of
interest.  Aside from knowing that the dynamics of interest
evolves according to a system of the form $d{\bf z}/dt = {\bf F}({\bf z})$, we
assume little or no additional knowledge of the system, and that
the available information is a limited-duration past time series
of the state evolution ${\bf z}(t)$. Nevertheless, we still desire
to deduce causal dependencies, where the meaning of causal is in terms of responses to perturbations as
defined above.  Since, as we will see, accomplishment of this
task, in principle, is not always possible, our approach will
 be to first propose a heuristic solution, and then
numerically test its validity.  The main message of this paper is
that our proposed procedure can be extremely effective for a very
large class of important problems. We will also delineate
situations where our procedure is expected to fail. We emphasize that, as our method is conceptually based on consideration of responses to perturbations, in our opinion, it provides a more direct test of what is commonly of interest when determining causality than do tests based on prediction impact, correlation, or entropy metrics.

Furthermore, although the setting motivating our procedure is for
deterministic systems, $d{\bf z}/dt = {\bf F}({\bf z})$, we will
also investigate performance of our procedure in the presence of
both dynamical noise (i.e., noise added to the state evolution equation,
$d{\bf z}/dt = {\bf F}({\bf z})$) and observational noise (i.e.,
noise added to observations of ${\bf z}(t)$ used as training data
for the machine learning).  Both types of noise
are, in practice, invariably present.  An important result from our study is that the presence of {\it
dynamical} noise can very greatly enhance the accuracy and applicability of our
method (a
similar point has been made in Ref.\ \cite{r7} and Ref.\ \cite{r7b}), while observational noise degrades the ability to infer causal dependence.

To more precisely define causal dependence, we consider the effect of a perturbation on one variable on the other variables as follows. Taking the $j^{{\rm th}}$ component of $d{\bf z}/dt = {\bf F}({\bf
z})$, we have
\[ dz_j(t)/dt = F_j(z_1(t), \ldots, z_2(t), \ldots, z_M(t)), \]
for $j=1,2,\ldots,M$.  Perturbing $z_i(t)$ by $\delta z_i(t)$, we
obtain for small $\tau$, that the component of the orbit perturbation of $z_j$ at
time $(t + \tau)$ due to $\delta z_i$ is
\[
\delta z_j(t+\tau) = \tau \left\{ \frac{\partial F_j({\bf
z})}{\partial z_i} | _{{\bf z} = z(t)} \right\} \delta z_i(t) +
{\mathcal O}(\tau^2).
\]
We define the {\it Short Term Causal Dependence} (STCD) metric, $f_{ji}$, of $z_j$ on $z_i$
by
%Eq. 1
\begin{equation}
f_{ji} =\left \langle G\left( \frac{\partial F_j({\bf z})}{\partial z_i}\right )\right
\rangle ,
\end{equation}
where $\langle (\ldots) \rangle$ denotes a long time average of
the quantity $( \ldots )$ over an orbit, and the function $G$ is to be chosen in a situation-dependent manner. For example, later in this paper, we consider examples addressing Goal (i) (where we want to distinguish whether or not ${\partial F_j({\bf z})}/{\partial z_i}$  is always zero) for  which we use $G(q)=|q|$, while, when we consider an example addressing Goal (ii) and are concerned with the time-averaged signed value of the interaction strength, we then use $G(q)=q$. In either case, we view $f_{ji}$ as quantifying the
causal dependence of $z_j$ on $z_i$, and the key goal of this paper
will be to obtain and test a machine learning procedure for
estimating $f_{ji}$ from observations of the state evolution ${\bf
z}(t)$.  For future reference, we will henceforth denote our
machine learning estimate of $f_{ji}$ by $\hat{f}_{ji}$.  In the case of our Goal (i) experiments, where $G(q)=|q|$, we note
that $f_{ji}$ defined by (1) is an average of a non-negative
quantity and thus $f_{ji} \geq 0$, as will be our estimate,
$\hat{f}_{ji} \geq 0$.  Furthermore, for that case we will define STCD of $z_i$ on $z_j$ by the
condition, $f_{ji} > 0$, and, when using our machine learning
estimate $\hat{f}_{ji}$, we shall judge STCD to likely apply when
$\hat{f}_{ji} > \epsilon$ where we call $\epsilon > 0$ the {\it
discrimination threshold}. In the ideal case $(\hat{f}_{ji} =
f_{ji})$, the discrimination threshold $\epsilon$ can be set to
zero, but, in practice, due to error in our estimate, we consider
$\epsilon$ to be a suitably chosen positive number. We note that, in the
ideal case, $\epsilon = 0$ can be regarded as a test for whether
or not $F_j({\bf z})$ is independent of $z_i$.

As a demonstration of a situation for which the determination of
STCD from observations of the motion of ${\bf z}(t)$ on its
attractor is not possible, we note the case where the attractor is
a fixed point (a zero-dimensional attractor).  Here, the measured
available information is the $M$ numbers that are the coordinates
of the fixed point, and this information is clearly insufficient
for determining STCD.  As another problematic example, we note
that in certain cases one is interested in a dynamical system that
is a connected network of identical dynamical subsystems, and that
such a network system can exhibit exact synchronization of its
component subsystems\textsuperscript{\cite{r8}}(including cases where the subsystem
orbits are chaotic).  In the case where such a synchronized state is stable, observations of the individual
subsystems are indistinguishable, and it is then impossible, in
principle, for one to infer causal relationships between state
variables belonging to different subsystems. More generally, in addition to the above fixed point and synchronization examples, we note that the dimension of the tangent space at a given point ${\bf z}^{*}$ on the attractor is, at most, the smallest embedding dimension of the part of the attractor in a small neighborhood of ${\bf z}^{*}$. Thus the full $M \times M$ Jacobian of ${\bf F}({\bf z})$ at ${\bf z}^{*}$ cannot be precisely determined from data on the attractor when the local attractor embedding dimension at ${\bf z}^{*}$ is less than $M$, which is commonly the case. Thus these examples motivate the conjecture that to efficiently and accurately infer STCD, the orbital complexity of the dynamics should be large enough so as to encode the information that we seek. Note that these considerations of cases where inference of STCD is problematic do not apply to situations with dynamical noise, e.g., $d{\bf z}/dt = {\bf F}({\bf z})+({\text{noise}})$, as the addition of noise may be roughly thought of as introducing an infinite amount of orbital complexity. Alternatively, the addition of noise increases the embedding dimension of the data to that of the full state space, i.e., $M$.

\noindent{\bf III. USING RESERVOIR COMPUTING TO DETERMINE  STCD}

We base our considerations on a type of machine learning called
reservoir computing, originally put forward in Refs. \cite{r9} and \cite{r10} (for a review, see Ref. \cite{r11}). We assume that we can sample the time-series data ${\bf z}(t)$ from our system at regular time intervals of length $\tau$, so that we have a discrete set of observations $\{ {\bf z}(0), {\bf z}(\tau), {\bf z}(2\tau),...\}$.  To begin, we first describe a reservoir-computer-based machine learning procedure in which the reservoir computer is trained to give an output $\hat{\bf z}(t+\tau)$ in response to an $M$-dimensional input ${\bf z}(t)$ as illustrated in Fig. 1.

For our numerical tests we consider a specific
reservoir computer implementation (Fig. 1) in which the reservoir consists
of a network of $R \gg M$ nodes whose scalar states, $r_1(n\tau),
r_2(n\tau), \ldots, r_R(n\tau)$, are the components of the
$R$-dimensional vector ${\bf r}(n\tau)$.

The nodes interact dynamically with each other through an $R
\times R$ network adjacency matrix ${\bf A}$, and their evolution
is also influenced by coupling of the $M$-dimensional input ${\bf
z}(n \tau)$ to the individual nodes of the reservoir network by
the $M \times R$ input coupling matrix ${\bf W}_{in}$ according to
the neural-network-type of evolution equation (e.g., Refs. \cite{r11,r12,r13,r14,referee1,referee2}) 
% Eq. 2
\renewcommand{\theequation}{2}
\begin{equation}
{\bf r}((n+1)\tau) = \tanh ({\bf A}{\bf r}(n \tau) + {\bf
W}_{in}{\bf z}(n \tau)) ,
\end{equation}
where $\tanh ({\bf v})$ for a vector ${\bf v} =
(v_1,v_2,v_3,\ldots)^T$ is defined as $(\tanh v_1,\tanh v_2,\tanh
v_3,\ldots)^T$.  For proper operation of the reservoir computer,
it is important that Eq.\ (2) satisfy the `echo state
property'\textsuperscript{\cite{r9,r11,r12}} (in nonlinear dynamics this condition is
also know as `generalized synchronization'\textsuperscript{\cite{r15,r16,r17}}): given two
different initial reservoir states, ${\bf r}_{1*}$ and ${\bf
r}_{2*}$, for the same input time series of ${\bf z}$, the
difference between the two corresponding reservoir states converges to
zero as they evolve in time (that is, $|{\bf r}_1(t) - {\bf
r}_2(t)| \rightarrow 0$ as $t \rightarrow \infty$, implying that,
after a transient initial period, ${\bf r}(t)$ essentially depends
only on the past history of ${\bf z}$, ${\bf z}(t^\prime)$ for
$t^\prime \leq t$, and not on the initial condition for ${\bf
r}$).

Using measured input training data over a training interval of
length $T \tau$, which begins after the initial transient period
mentioned above, we use Eq. (2) to generate ${\bf r}(\tau), {\bf
r}(2\tau), \ldots,{\bf r}(T\tau)$. We also record and store these
determined values ${\bf r}(n \tau)$ along with the corresponding
inputs, ${\bf z}(n \tau)$ that created them.   The matrices ${\bf
A}$ and ${\bf W}_{in}$ are regarded as fixed and are typically
chosen randomly.  In contrast, the $R \times M$ output coupling
matrix ${\bf W}_{out}$, shown in Fig.\ 1, is regarded as an
adjustable linear mapping from the reservoir states ${\bf r}$ to
an $M$-dimensional output vector $\hat{{\bf z}}$,
%Eq. 3
\renewcommand{\theequation}{3}
\begin{equation}
\hat{{\bf z}} ((n+1)\tau) = {\bf W}_{out} {\bf r}((n+1)\tau).
\end{equation}
`Training' of the machine learning reservoir computer then
consists of choosing the $RM$ adjustable matrix elements
(`weights') of ${\bf W}_{out}$ so as to make $\hat{{\bf
z}}(n\tau)$ a very good approximation to ${\bf z}(n
\tau)$ over the time duration $(\tau,2\tau,\ldots,T\tau)$ of the
training data.  This is done by minimization with respect to ${\bf
W}_{out}$ of the quantity,\\ 
$\left\{ \sum^T_{n=1} \parallel {\bf
z}(n \tau) - {\bf W}_{out} {\bf r}(n\tau) \parallel^2\right\} +
\beta \parallel {\bf W}_{out} \parallel^2$.  Here $\beta \parallel
{\bf W}_{out}
\parallel^2$, with $\beta$ small, is a `ridge' regularization
term\textsuperscript{\cite{r18}} added to prevent overfitting and $({\bf r}(n\tau),
{\bf z}(n\tau))$ are the previously recorded and stored training
data.  In general, $R \gg M$ is required in order to obtain a good
fit of $\hat{{\bf z}}$ to ${\bf z}(t)$.  For illustrative purposes we now consider the ideal
case where $\hat{{\bf z}} = {\bf z}$ (i.e., the training perfectly
achieves its goal).

 For the purpose of estimating
STCD, we now wish to eliminate the quantity ${\bf r}$ from the
basic reservoir computer system (Eqs.\ (2) and (3)) to obtain an
evolution equation solely for the state variable ${\bf z}$.  To do
this, we would like to solve (3) for ${\bf r}$ in
terms of ${\bf z}$.  However, since $R$, the dimension of ${\bf
r}$, is much larger than $M$, the dimension of ${\bf z}$, there
are typically an infinite number of solutions of (3) for ${\bf
r}$. To proceed, we hypothesize that it may be useful to eliminate
${\bf r}$ by choosing it to be the solution of (3) with the
smallest $L_2$ norm.  This condition defines the so-called
Moore-Penrose inverse\textsuperscript{\cite{r19}} of ${\bf W}_{out}$, which we denote
$\hat{\bf W}^{-1}_{out}$; i.e., the minimum  $L_2$ norm solution
for ${\bf r}$ is written ${\bf r} = \hat{\bf W}^{-1}_{out} {\bf
z}$. 
We emphasize that $\hat{\bf W}^{-1}_{out}{\bf z}$ is not necessarily expected
to give the correct ${\bf r}$ obtained by solving the system, (2)
and (3).  However, from numerical results to follow, our choice
will be supported by the fact that it often yields very useful
estimates of $f_{ji}$.

Now applying ${\bf W}_{out}$ to both sides of Eq.\ (2) and,
employing ${\bf r} =\hat {\bf W}^{-1}_{out} {\bf z}$ to eliminate
${\bf r}(n\tau)$ from the argument of the $\tanh$ function in Eq.\
(2), we obtain a surrogate time -$\tau$ map for the evolution of
${\bf z}$, ${\bf z}((n+1)\tau) = {\bf H}[{\bf z}(n\tau)]$, where
${\bf H}({\bf z}) = {\bf W}_{out} \tanh [({\bf A}\hat{\bf
W}^{-1}_{out} + {\bf W}_{in}){\bf z} ]$ .  Here we note that we do not claim that this map in itself can be used for time-series prediction in place of Eqs. (2) and (3), which were commonly used in previous works (e.g., Refs. \cite{r12,r13,r14,referee1,referee2}) . Rather, we use it as a symbolic represention of the result obtained after eliminating the reservoir state vector ${\bf r}$ from Eqs. (2) and (3). In particular, the prediction recipe using Eqs. (2) and (3) is always unique and well-defined, in contrast to the above map, where $ {\bf W}^{-1}_{out}$ is clearly non-unique. So, we use this map only for causality estimation purposes, as described below. Differentiating ${\bf
H}({\bf z})$ with respect to $z_i$, we have
%Eq. 5
\renewcommand{\theequation}{4}
\begin{equation}
\frac{\partial F_j({\bf z})}{\partial z_i} = \tau^{-1} \left[
\frac{\partial H_j({\bf z})}{\partial z_i} - \delta_{ij} \right] ,
\end{equation}
where $\delta_{ij}$ is the Kronecker delta, and we propose to use Eqs. (1) and (4) to determine STCD.

In our numerical experiments, the number of training time steps is $T=6\times 10^4$ for Figs. 2, 3 and $T=2\times 10^4$  for Fig. 4. In each case, the actual training data is obtained after discarding a transient part of $2\times 10^4$ time steps and the
reservoir system sampling time is $\tau=0.02$.  The elements of
the input matrix ${\bf W}_{in}$ are randomly chosen in the
interval $[-0.1,0.1]$.  The reservoir is a sparse random network of
$R=5000$ nodes for Figs. 2, 3 and of $R=1000$ nodes for Fig. 4. In each case the average number of incoming links per
node is $3$.  Each nonzero element of the reservoir adjacency
matric ${\bf A}$ is randomly chosen from the interval $[-a,a]$,
and $a>0$ is then adjusted so that the maximum magnitude
eigenvalue of ${\bf A}$ is $0.9$.  The regularization parameter is
$\beta =10^{-4}$. These parameters are adapted from Ref. \cite{r14}. The average indicated in Eq. (1) is over $1000$ time steps. The chosen time step $\tau$ is sufficiently
small compared to the timescale over which ${\bf z}(t)$ evolves 
that the discrete time series ${\bf z}(n \tau)$ is a good
representation of the continuous variation of ${\bf z}(t)$. 

%Figure 1
\begin{figure}[h!]
\begin{center}
\includegraphics[width=5in]{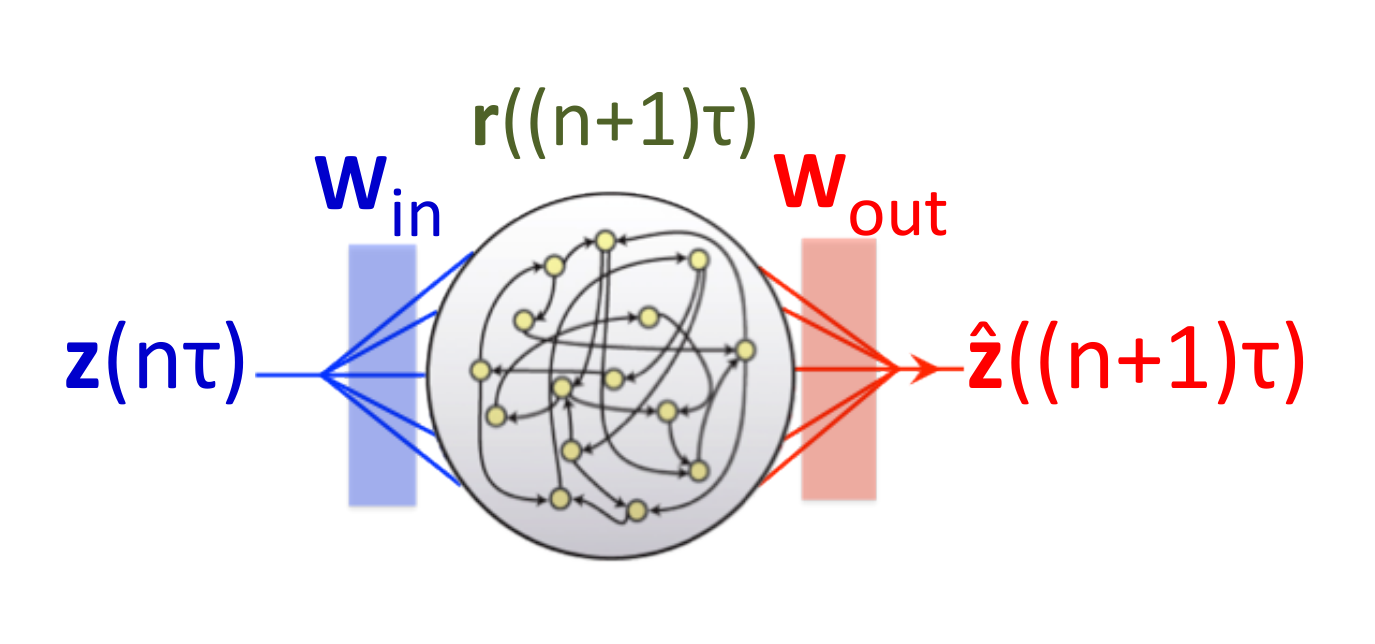}
\end{center}
\renewcommand{\baselinestretch}{1}
\small\normalsize
\begin{quote}
\caption{Schematic of the reservoir computing architecture used in this work.  The input-to-reservoir coupling matrix ${\bf W}_{in}$ couples the input time series for the vector ${\bf z}$ to the reservoir state vector ${\bf r}$. The reservoir-to-output coupling matrix ${\bf W}_{out}$ generates the output vector $\hat{\bf z}$ from the reservoir. $\hat{\bf z}$ is found to be a good estimate of ${\bf z}$ after training.}
\end{quote}
\end{figure}
\renewcommand{\baselinestretch}{2}
\small\normalsize
Although we use a specific reservoir computing
implementation, we expect that, with suitable
modifications, our approach can be
adapted to `deep' types of machine learning\textsuperscript{\cite{r2}}, as well as to other
implementations of reservoir computing\textsuperscript{\cite{r21,r22,r20,r23}}, (notably implementations involving photonics\textsuperscript{\cite{r21}}, electronics\textsuperscript{\cite{r20}} and field programmable gate arrays(FPGAs)\textsuperscript{\cite{r22}}).

\noindent {\bf IV. TESTS OF MACHINE LEARNING INFERENCE OF STCD}

In order to evaluate the effectiveness of our proposed method, we
introduce mathematical model test systems that we use as proxies
for the unknown system of interest for whose state variables we wish to determine STCD.  We next use the test systems to generate simulated
training data from which we determine STCD by our ML technique. We
then assess the performance of the technique by the correctness of
its results determined from the known properties of the test
systems.

We first consider examples addressing our Goal (i) ($G(q)=|q|$ in Eq. (1)), and for our simulation test systems, we consider the case of a network
of $N$ nodes and $L$ links, where each node is a classical Lorenz
system\textsuperscript{\cite{r24}} with heterogeneity from node to node, additive
dynamical noise, and internode coupling,\useshortskip
%Eq.7
\renewcommand{\theequation}{5}
\begin{equation}
dx_k/dt = - 10[x_k-y_k+c \sum^N_{l=1} a^{(x,y)}_{kl} (y_l-y_k)] +
\sigma_{\text{Dyn}} n_{kx}(t) ,
\end{equation}
%Eq.7

\useshortskip
\renewcommand{\theequation}{6}
\begin{equation}
dy_k/dt = 28 (1 + h_k)x_k-y_k-x_kz_k + \sigma_{\text{Dyn}} n_{ky}(t) ,\\
\end{equation}

\useshortskip
%Eq.7
\renewcommand{\theequation}{7}
\begin{equation}
dz_k/dt = - (8/3)z_k + x_ky_k + \sigma_{\text{Dyn}} n_{kz}(t) .
\end{equation}

The state space dimension of this system is $M=3N$.  The coupling
of the $N$ nodes is taken to be only from the $y$ variable of one
node to the $x$ variable of another node with coupling
constant $c$, and $a^{(x,y)}_{kl}$ is either 1 or 0 depending on
whether or not there is a link from $l$ to $k$.  The adjacency
matrix $a^{(x,y)}_{kl}$ of our Lorenz network (not to be confused
with the adjacency matrix ${\bf A}$ of the reservoir) is
constructed by placing directed links between $L$ distinct
randomly chosen node pairs.  For each node $k$, $h_k$ is randomly
chosen in the interval $[-h,+h]$, and we call $h$ the heterogeneity parameter.  Independent white noise terms of equal
variance $\sigma_{\text{Dyn}}^2$ are added to the left-hand sides of the
equations for $dx/dt$, $dy/dt$ and $dz/dt$, where, for example,
$\left \langle n_{kx}(t) n_{k'x}(t')\right \rangle = 2 \delta_{kk'} \delta(t-t')$.  For
$\sigma=c=h=0$, each node obeys the classical chaotic Lorenz equation with
the parameter values originally studied by Lorenz\textsuperscript{\cite{r24}}.  Furthermore, denoting the right-hand
side of Eq.\ (5) by $F_{xk}$, we have $\partial F_{xk}/\partial
y_l = 10 c$ or 0, depending on whether there is, or is not, a link
from $y_l$ to $x_k$.

Since in this case, the derivative $\partial F_{xk}/\partial y_l$ is time independent, $\left \langle \left|
\partial F_{xk}/\partial y_l\right |\right \rangle$ is also either $10c$ or 0, and, adopting
the notation $f^{(x,y)}_{kl} = \left \langle \left|
\partial F_{xk}/\partial y_l\right |\right \rangle$,
we denote its machine learning estimate by our previously
described procedure by $\hat{f}^{(x,y)}_{kl}$.  For a reasonably
large network, the number $N^2-N$ of ordered node pairs $(k,l)$ of distinct nodes is
large, and we consequently have many values of
$\hat{f}^{(x,y)}_{kl}$. Bayesian techniques (see Ref. \cite{r25} and references therein) can be applied to such data to obtain an estimate $\hat{L}$ for the total number of links $L$, and one can then set the value of
$\epsilon$ so that there are $\hat{L}$ values of
$\hat{f}^{(x,y)}_{kl}$ that are greater than $\epsilon$.  Less formally, we find that making a histogram of the values of $\hat{f}^{(x,y)}_{kl}$ often reveals a peak at zero and another peak at a higher positive value with a large gap or discernible minimum in between. One can then estimate $\epsilon$ by a value in the gap or by the location of the minimum between the peaks, respectively. For
simplicity, in our illustrative numerical simulations to follow we
assume that $L$ is known (approximately equivalent to the case that $L$ is unknown but that a very good estimate ($\hat{L}$) has been obtained).

{\it Example 1: A heterogeneous noiseless case}.  We consider the
parameter set $c=0.3$, $h=0.06$, $\sigma_{\text{Dyn}}=\sigma_{\text{Obs}}=0$, $N=20$, and we vary
the number of links $L$.  Figure 2(a) (for $L=50$) and (b) (for
$L=100$) each show an array of 20$\times$20 boxes where each of the boxes represents an ordered node pair $(k,l)$ of the 20-node network,
and the boxes have been colored (see Table 1) according to whether the results
for our procedure predict a link from $l$ to $k$ (``positive'') or not
(``negative''), and whether the prediction is correct (``true'')
or wrong (``false'').
\begin{table}[h]
\centering
\begin{tabular}{|m{4.5cm}|m{1cm}|}
\hline
{\bf TP (True Positive)}  & Black Square \\ \hline 
{\bf TN (True Negative)}  & White Square \\ \hline
{\bf FP (False Positive)}  & Blue Square\\ \hline
{\bf FN (False Negative)} & Red Square \\  \hline
\end{tabular}
\caption{Color-coding scheme for Figs. 2 and 3.}

\end{table}

We see that for a typical case with $L=50$ (Fig. 2(a)) {\it all} the boxes have been correctly labeled,
corresponding to all boxes being either black or white.  In
contrast to this perfect result at $L=50$, at $L=100$ (Fig. 2(b)) the method
fails terribly, and the fraction of correct inferences is small.  In
fact, we find excellent performance for $L \leq 50$, but that, as
$L$ increases past 50, the performance of our method degrades
markedly. This is shown in Fig.\ 2(c) where we give plots of the number of false positives (FP) normalized to the expected value of FP that would result if $L$ links were randomly assigned to the $N^2-N=380$ node pairs $(k,l)$. (We denote this normalization $\left \langle{\text{FP}}\right \rangle_{R}$; it is given by $\left \langle{\text{FP}}\right \rangle_{R}={L(380-L)}/{380}$.)  Note that, with this normalization, for the different heterogeneities plotted in Fig.\ 2(c), the curves are similar, and that they all begin increasing at around $L=60$ and 
$\text{FP}/\left \langle{\text{FP}}\right \rangle_{R}$ becomes nearly $1$ (i.e., inference no better than random) past $ L\sim100$.  In our earlier discussion we have
conjectured that, for inference of STCD to be possible, the
orbital complexity should not be too small. To test this
conjecture we have calculated the information dimension $D_{INFO}$
of the network system attractor corresponding to the parameters,
$c=0.3$, $h=0$, $\sigma=0$, $N=20$, as a function of $L$.  We
do this by calculating the Lyapunov exponents of the system Eqs.\
(5)-(7), and then applying the Kaplan-Yorke formula for $D_{INFO}$ in
terms of the calculated Lyapunov exponents.\textsuperscript{\cite{r26,r27}}  The result
is shown in Fig.\ 2(d), where we see that $D_{INFO}$ decreases
with increasing $L$.  Regarding $D_{INFO}$ as a measure of the
orbital complexity, this is consistent with our expectation that the
ability to infer STCD will be lost if the orbital complexity of
the dynamics is too small. As we next show, the above negative result for $L$ increasing past about $60$ does not apply when even small dynamical noise is present.

%Figure 2
\begin{figure}[h!]
\begin{center}
\includegraphics[width=3in]{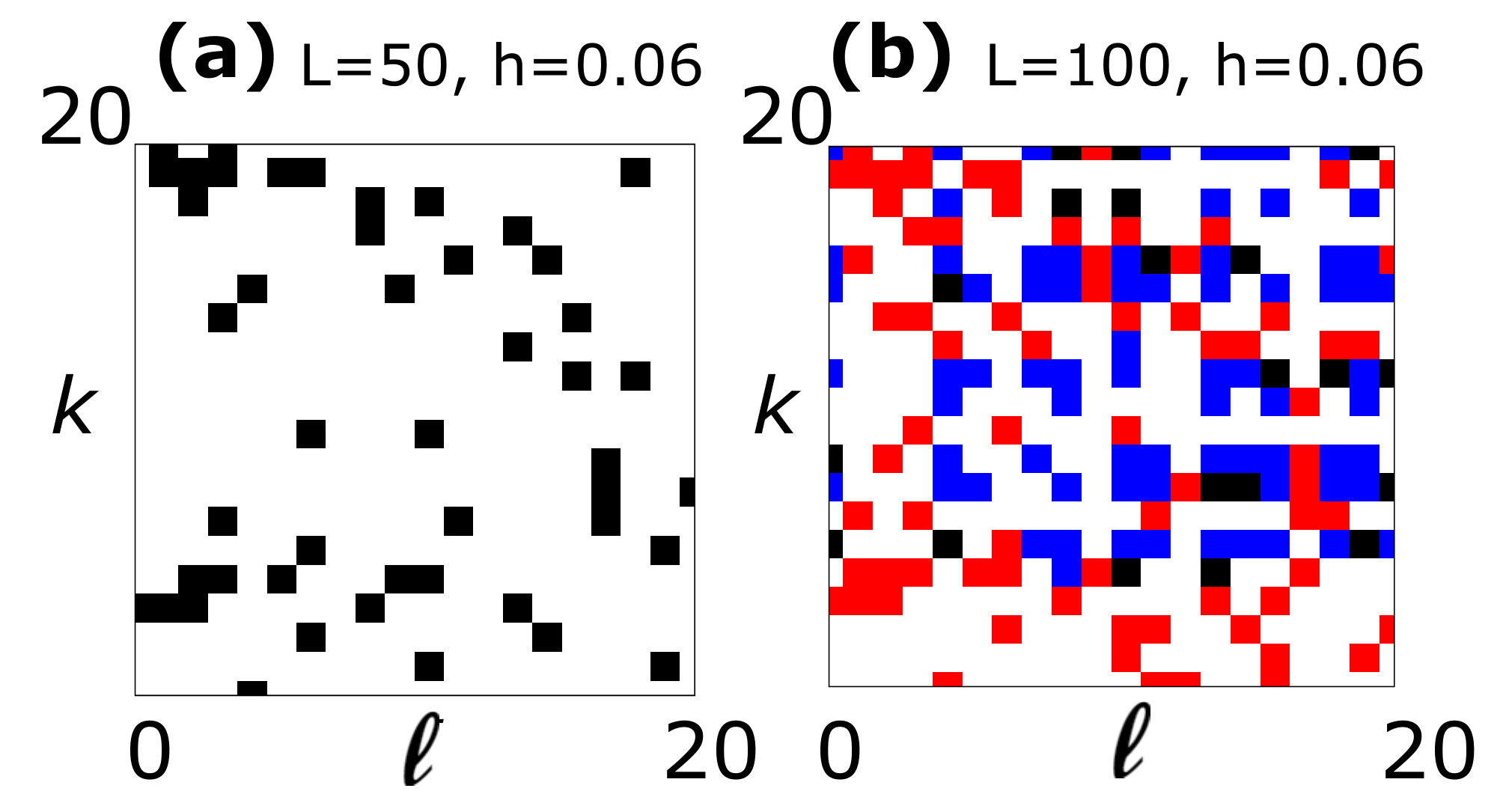}\\
\includegraphics[width=3in]{fig2b.pdf}
\end{center}
\renewcommand{\baselinestretch}{1}
\small\normalsize
\begin{quote}
\caption{Results of Experiment 1 (noiseless case). Panels (a) and (b) show the results of link inferences for two noiseless
cases for $L=50$ links and $L=100$ links.  The inference is perfect
in (a), but is very bad in (b). (c) $\text{FP}/\left \langle{\text{FP}}\right \rangle_{R}$ versus $L$ for $h=0,0.06,0.15$ averaged over $100$ random realizations of the system and the reservoir adjacency matrix. (d) The orbital complexity as
measured by the attractor information dimension $D_{INFO}$
decreases with increasing $L$. Note that at each value of $L$, we compute the $D_{INFO}$ for 10 random realizations of a network with $L$ links with $h=0$. The Kaplan-Yorke dimension is then averaged over all network realizations and the resulting plot is further smoothed by applying a moving average filter.}
\end{quote}
\end{figure}
\renewcommand{\baselinestretch}{2}
\small\normalsize

{\it Example 2:  The effects of dynamical and observational
noise.} We first consider the effect of dynamical noise of
variance $\sigma_{\text{Dyn}}^2$ for the parameters $h=0$ (homogeneous), $c=0.3$,
$N=20$, $L=200$.  Results (similar in style to  Figs.\ 2(a) and
2(b)) are shown in  Figs.\ 3(a), 3(b), and 3(c).  For extremely
low dynamical noise variance, $\sigma_{\text{Dyn}}^2=10^{-9}$ (Fig.\ 3(a)), the result
is essentially the same as for zero noise, and about one quarter of
the boxes are classified TP, TN, FP, and FN each (since there are
200 links and 400 boxes, this is no better than random
assignment). As the noise variance is increased to $\sigma_{\text{Dyn}}^2 =
10^{-7.5}$ (Fig.\ 3(b)), the results become better, with a fraction 0.75 of
the boxes either TP or TF (as opposed to 0.52 for Fig. 3(a)).  Upon further increase of the dynamical noise variance to the
still small value of $\sigma_{\text{Dyn}}^2=10^{-6}$ (Fig.\ 3(c)), the results typically
become perfect or nearly perfect.  Furthermore, excellent results, similar to
those for $\sigma_{\text{Dyn}}^2=10^{-6}$, continue to apply for larger $\sigma_{\text{Dyn}}^2$.
This is shown by the red curve in Fig. 3(f) which shows $\text{FP}/\left \langle{\text{FP}}\right \rangle_{R}$ versus $\sigma_{\text{Dyn}}^2$ ($N=20; L=200$).  Importantly, we also note that our normalization of FP by ${\left \langle{\text{FP}}\right \rangle_{R}}$ essentially makes the red curve $L$-independent over the range we have tested, $50 \leq L \leq 200$. Our interpretation of
this dynamical-noise-mediated strong enhancement of our ability to
correctly infer links is that the dynamical noise allows the orbit
to explore the state space dynamics off the orbit's attractor and
that the machine learning is able to make appropriate good use of
the information it thus gains.
%Figure 3
\begin{figure}[h!]
\begin{center}
\includegraphics[width=5in]{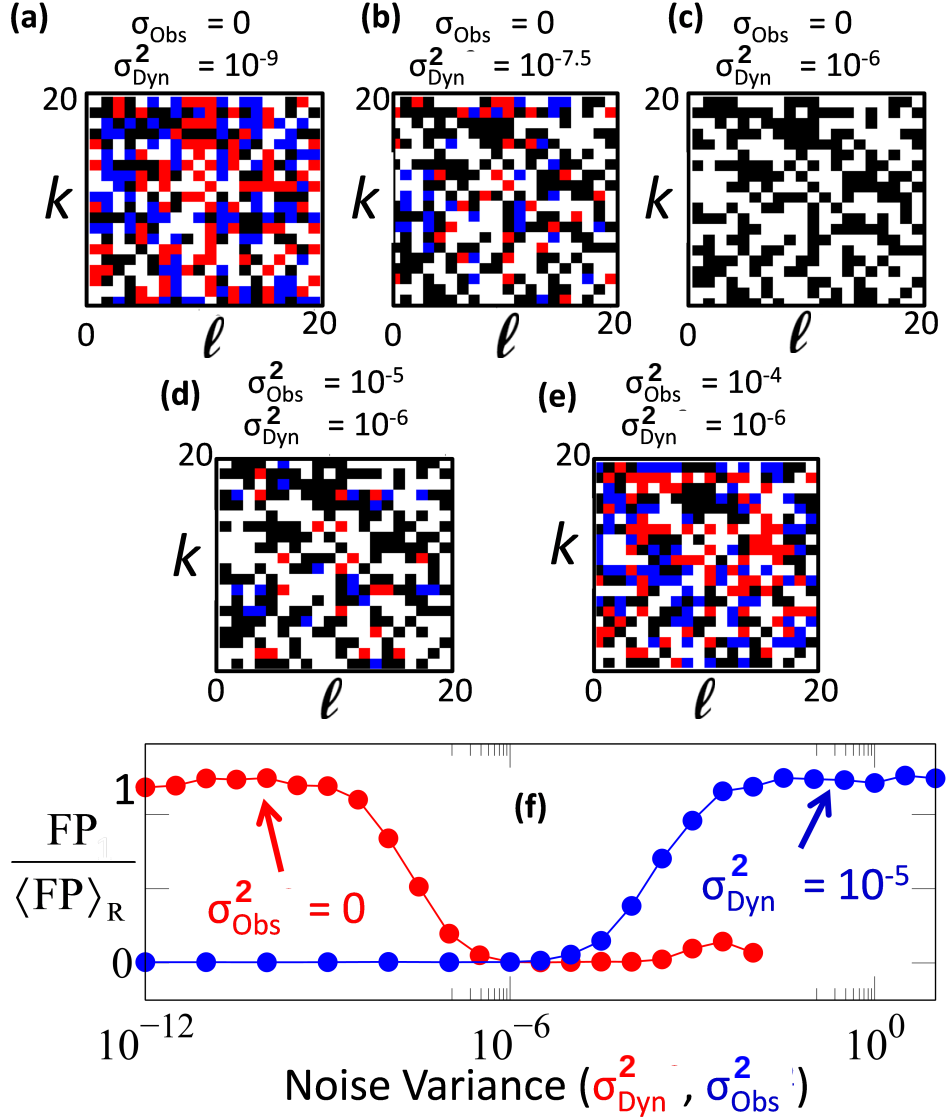}
\end{center}
\renewcommand{\baselinestretch}{1}
\small\normalsize
\begin{quote}
\caption{The effect of noise on STCD inference. Panels (a), (b), and (c) shows the effect of increasing the dynamical
noise variance $\sigma_{\text{Dyn}}^2$ is to greatly enhance the effectiveness of
link identification even at the rather low noise level of
$\sigma_{\text{Dyn}}^2=10^{-6}$.  In contrast, as shown in panels (d), (e), and (f),
starting with the situation (c) and increasing the observational
noise variance $\sigma_{\text{Obs}}^2$ degrades link identification. $L=200, h=0$ for all the subfigures here.}
\end{quote}
\end{figure}
\renewcommand{\baselinestretch}{2}
\small\normalsize

We now turn to the effect of observational noise by replacing the
machine learning time series training data formerly used,
$[x_k(n\tau), y_k(n\tau), z_k(n\tau)]$, by $[x_k(n\tau) +
\hat{\sigma}_{\text{Obs}} \hat{n}_{kx}(n\tau), y_k(n\tau) +
\hat{\sigma}_{\text{Obs}} \hat{n}_{ky}(n\tau),z_k(n\tau) +
\hat{\sigma}_{\text{Obs}}\hat{n}_k(n\tau)]$, where the parameter ${\sigma}_{\text{Obs}}^2$
is the observational noise variance and the $\hat{n}_{kx},
\hat{n}_{ky},\hat{n}_{kz}$ are independent Gaussian random variables with,
e.g., $\left \langle\hat{n}_{kx}(n\tau) \hat{n}_{k'x}(n'\tau)\right \rangle = 2
\delta_{kk'}\delta_{nn'}$.  The blue curve in Fig. 3(f) shows the
effect of adding observational noise of variance ${\sigma}_{\text{Obs}}^2$ on
top of dynamical noise for the situation $\sigma_{\text{Dyn}}^2=10^{-5}$ of Fig.\ 3(c).  We see from Figs. 3(d)-(f) that, when
${\sigma}_{\text{Obs}}^2$ is below about $10^{-5}$, it is too small to have much effect,
but, as ${\sigma}_{\text{Obs}}^2$ is increased above $10^{-5}$, the observational
noise has an increasing deleterious effect on link inference.  This
negative effect of observational noise is to be expected, since
inference of characteristics of the unknown system is necessarily
based on the part of the signal that is influenced by the dynamics
of the unknown system, which the observational noise tends to
obscure.

{\it Example 3:  Inferring Continuous Valued Dependence
Strengths.}  We now wish to  address Goal (ii) (for which we take $G(q)=q$ in Eq. (1)) and we, accordingly, consider the case where $f^{(x,y)}_{kl}$ for each
$(k,l)$ takes on a value in a continuous range (rather than the case of {\it
Examples 1 and 2} where $f^{(x,y)}_{kl}$ is either $10 c$ or zero
for all $(k,l)$).  For this purpose we replace Eq.\ (5) by
%Eq. 10
\renewcommand{\theequation}{8}
\begin{equation}
dx_k/dt = - 10(x_k - y_k)+\sum_l f^{(x,y)}_{kl} y_l ,
\end{equation}
and consider Eqs.\ (6), (7), and (8) as our new test system, with
$h=0.9$, ${\sigma}_{\text{Dyn}}^2={\sigma}_{\text{Obs}}^2=0$, and $N=100$ nodes (corresponding to
$100\times 100 = 10^4$ possible connection strength values).  We
choose the connection strength values as follows.  A
photographic portrait of Edward N.\ Lorenz is divided up into
$100\times 100 = 10^4$ boxes and, by using a shading scale from dark
(coded as +10) to light (coded as -5), Fig.\ 4(a) is obtained, with
the shading scale given to the right of Fig. 4(b).  Setting
$f^{(x,y)}_{kl}$ equal to the color scale value of box $(k,l)$, we
next numerically solve Eqs. (6), (7), and (8).  We then use this
orbit as the training data for input to our ML determination of
causal strength dependence, $\hat{f}^{(x,y)}_{kl}$, and employing
the same shading scale, use the thus determined
values of $\hat{f}^{(x,y)}_{kl}$ to reconstruct the original
portrait, as shown in Fig.\ 4(b).  We see that, although the
reproduction is not exact, the overall picture is still clearly
recognizable, indicating the effectiveness of the method for Goal (ii).
 For a more quantitative comparison of the actual and the estimated Jacobian elements, we calculate the normalized Frobenius norm of their difference matrix $f^{(x,y)}-\hat{f}^{(x,y)}$. We first apply upper and lower cut-offs equal to 10 and -5.5 respectively to $\hat{f}^{(x,y)}$, in order to eliminate some extreme values. Then we calculate the ratio
\renewcommand{\theequation}{9}
\begin{equation}
\delta =\frac{\left|\left|f^{(x,y)}-\hat{f}^{(x,y)} \right|\right|_{F}}{\left \langle \left|\left|f^{(x,y)}-\tilde{f}^{(x,y)} \right|\right|_{F}\right \rangle},
\end{equation}
where $\left|\left|M\right|\right|_{F}=\sqrt{{\text {Trace}}(M^{\dagger}M)}=\sqrt{\sum\limits_{i,j}\left|M_{ij}\right|^2}$ is the Frobenius norm of the matrix $M$. Here $\tilde{f}^{(x,y)}$ denotes a matrix constructed by randomly permuting the elements of the matrix ${f}^{(x,y)}$, and the angled brackets denote an average over such random permutations. So this ratio compares the total error in the inferred Jacobian with the variation in the original matrix elements of ${f}^{(x,y)}$. For example, for a perfect estimation of ${f}^{(x,y)}$, we will have $\delta=0$. In contrast, $\delta=1$ means that the prediction error is equal to the average error when the elements of ${f}^{(x,y)}$ are randomized. For the example shown in Fig.\ 4, we find that $\delta$ is approximately equal to 0.37.

%Figure 4
\begin{figure}[h!]
\begin{center}
\includegraphics[width=5in]{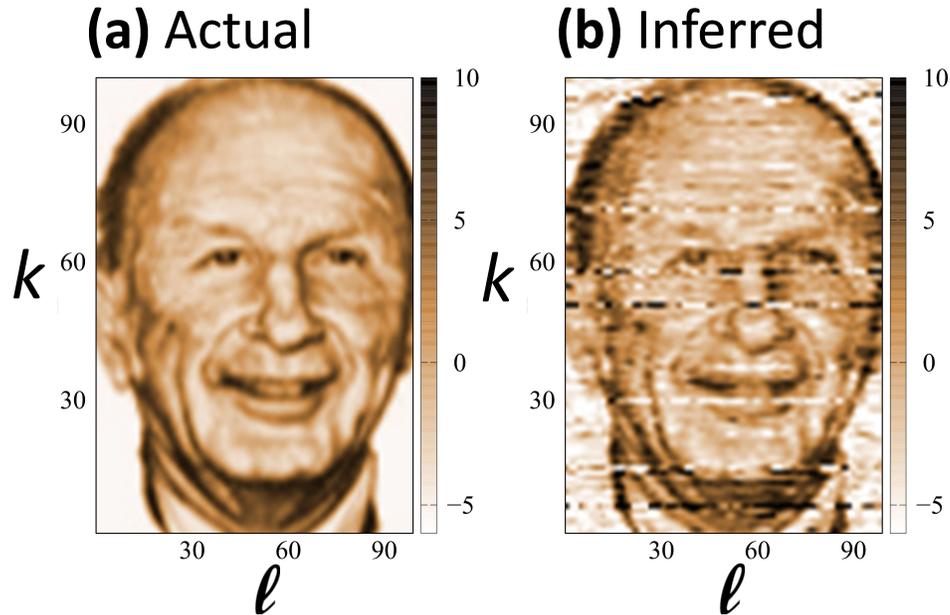}
\end{center}
\renewcommand{\baselinestretch}{1}
\small\normalsize
\begin{quote}
\caption{Results of Experiment 3. Panel (a) shows a 100$\times$100 pixelated, shade-coded portrait of
Edward N. Lorenz; (b) reconstruction of (a) by our ML link
inference technique. Note that, in (b), we plot all the values greater than or equal to $10$ as black and all the values less than or equal to $-5.5$ as white.}
\end{quote}
\end{figure}
\renewcommand{\baselinestretch}{2}
\small\normalsize

{\bf V. DISCUSSION}

In this paper, we have formulated and tested a new, highly effective, machine-learning-based approach for inferring causal dependencies of state variables of an unknown system from time series observations of these state variables. A key finding is that the effectiveness of our approach is greatly enhanced in the presence of sufficient dynamical noise, provided that the deleterious effect of observational noise is not too great. {\it The competition between the opposing effects of these two types of noise will likely be the essential key factor determining the success or failure of causality inference in many of the most important situations of interest (e.g., in neuroscience and genomics).} Much work remains to be done to more fully address the utility of our method. In particular, further numerical tests on diverse systems, and, especially, experimental studies in real world applications, will ultimately determine the circumstances under which the method developed here will be useful.

{\bf ACKNOWLEDGMENTS}

This work was supported by the U. S. National Science Foundation (Grant DMS 1813027). The authors acknowledge useful discussion with Sarthak Chandra, Amitabha Sen, Woodrow Shew, Nuno Martins, Adrian Papamarcou,
 Erik Bollt, and especially Brian Hunt.


{\bf REFERENCES}

\begin{enumerate}
\bibitem{r1} R.\ Feynman, {\it The Character of Physical Law} (MIT Press,
Cambridge, 1965).

\bibitem{r2} I.\ Goodfellow, Y.\  Bengio and A.\  Courville, {\it Deep
Learning} (MIT Press, Cambridge, 2016).

\bibitem{r3} C.W.J.\ Granger, ``Investigating Causal Relations by
Econometric Methods and Cross-Spectral Methods,'' Econometrica
{\bf 37}, 424-430 (1969).

\bibitem{r4} M.\ Ding, Y.\ Chen. and S.\ L.\ Bressler, ``Granger
Causality: Basic Theory and Applications to Neuroscience,'' in
{\it Handbook of Time Series Analysis}, pp.\ 437-460 (Wiley-VCH,
2006).

\bibitem{r4b} C. Sima, J. Hua, and S. Jung, ``Inference of Gene Regulatory Networks Using Time-Series Data: A Survey'', Curr Genomics. {\bf 10}(6): 416–429 (2009).

\bibitem{r5a} J. F. Donges, Y. Zou, N. Marwan, and J. Kurths, ``Complex networks in climate dynamics'', Eur. Phys. J. Spec. Top. {\bf 174}: 157 (2009).

\bibitem{r5} W.\ L.\ Ku, G.\ Duggal, Y.\ Li, M.\ Girvan, and E.\ Ott,
``Interpreting Patterns of Gene Expression:   Signatures of
Coregulation, the Data Processing Inequality, and Triplet
Motifs,'' PLoS One {\bf 7}, e31969 (2012).

\bibitem{r5new} J. Ren, W.-X. Wang, B. Li, and Y.-C. Lai, ``Noise Bridges Dynamical Correlation and Topology in Coupled Oscillator Networks'',
Phys. Rev. Lett. {\bf 104}, 058701 (2010).

\bibitem{r5new2} Z. Levnajic and A. Pikovsky, ``Untangling complex dynamical systems via derivative-variable correlations'', Sci. Rep. {\bf 4}, 5030 (2014); M. G. Leguia, R. G. Andrzejak, and Z. Levnajic, ``Evolutionary optimization of network reconstruction from derivative-variable correlations'',  J. Phys. A: Math. Theor. {\bf 50}, 334001 (2017).

\bibitem{r6} T.\ Schreiber, ``Measuring Information Transfer,'' Phys.\
Rev.\ Lett.\ {\bf 85}, 461-464.

\bibitem{r7a} J. Sun, and E. M. Bollt, ``Causation entropy identifies indirect influences, dominance of neighbors and anticipatory couplings'',   Physica D {\bf 267}, 49-57 (2014).

\bibitem{r7c} E. J. Molinelli, A. Korkut, W. Wang, M. L. Miller, N. P. Gauthier, X. Jing, P. Kaushik, Q. He, G. Mills, D. B. Solit, C. A. Pratilas, M. Weigt, A. Braunstein, A. Pagnani, R. Zecchina, and C. Sander, ``Perturbation Biology: Inferring Signaling Networks in Cellular Systems'', PLoS Comput Biol {\bf 9} (12): e1003290 (2013).

\bibitem{r7new} M. Timme, ``Revealing Network Connectivity from Response Dynamics'', M. Timme, 
Phys. Rev. Lett. {\bf 98}, 224101 (2007).

\bibitem{r7} M.\ J.\ Panaggio, M.-V.\ Ciocanel, L.\ Lazarus, C.\ M.\
Topaz, and B.\ Xu, ``Model Reconstruction from Temporal Data for
Coupled Oscillator Networks," 
arXiv: \\ 1905.01408v1, 4 May 2019.

\bibitem{r7b} M. G. Leguia, C. G. B. Martinez, I. Malvestio, A. T. Campo, R. Rocamora, Z. Levnajic, and R. G. Andrzejak, ``Inferring directed networks using a rank-based connectivity measure'',
 Phys. Rev. E {\bf 99}, 012319 (2019).
 
\bibitem{r7g} T. Stankovski, T. Pereira, P. V. E. McClintock, A. Stefanovska, ``Coupling functions: Universal insights into dynamical interaction mechanisms `',  Rev. Mod. Phys. {\bf 89}, 045001 (2017).

\bibitem{rnew4} S. G. Shandilya and M. Timme, ``Inferring network topology from complex dynamics'',
New J. Phys. {\bf 13}, 13004 (2011).

\bibitem{r7d} S. Leng, Z. Xu, and  H. Ma, ``Reconstructing directional causal networks with random forest: Causality meeting machine learning'', Chaos {\bf 29}, 093130 (2019).

\bibitem{r7e} R.-M. Cao, S.-Y. Liu, and  X.-K. Xua, ``Network embedding for link prediction: The pitfall and improvement'', Chaos {\bf 29}, 103102 (2019).

\bibitem{r7f} M. G. Leguia,  Z. Levnajic,  L. Todorovski, and  B. Zenko, ``Reconstructing dynamical networks via feature ranking'', Chaos {\bf 29}, 093107 (2019).

\bibitem{r12} H.\ Jaeger and H.\ Haas, ``Harnessing Nonlinearity:
Predicting Chaotic Systems and Saving Energy in Wireless
Communication,'' Science {\bf 304}, 78-80 (2004).

\bibitem{r13} J.\ Pathak, B.\ Hunt, M.\ Girvan, Z.\ Lu, and  E.\ Ott,
``Model-Free Prediction of Large Spatiotemporally Chaotic Systems
from Data: A Reservoir  Computing Approach,'' Phys.\ Rev.\ Lett.\
{\bf 120}, 024102 (2018).

\bibitem{r14} Z.\ Lu, J.\ Pathak, B.\ Hunt, M.\ Girvan, R.\ Brockett, and
E.\ Ott, ``Reservoir Observers:  Model-Free Inference of
Unmeasured Variables in Chaotic Systems,'' Chaos {\bf 27}, 041102
(2017).

\bibitem{r21} L. Larger, A. Baylon-Fuentes, R. Martinenghi, V. S. Udaltsov, Y. K. Chembo, and M. Jacquot, ``High-Speed Photonic Reservoir Computing Using a Time-Delay-Based Architecture: Million Words per Second Classification'', Phys. Rev. X {\bf 7}, 011015 (2017).

\bibitem{r22} B. Schrauwen, M. D'Haene, D. Verstraeten, and J. Van Campenhout, ``Compact hardware liquid state machines on FPGA for real-time speech recognition'', Neural Networks {\bf 21}, 511-523 (2008).


\bibitem{r8} L.\ M.\ Pecora and T.\ L.\ Carroll, ``Master Stability
Function for Synchronized Chaotic Systems,'' Phys.\ Rev.\ Lett.\
{\bf 80}, 2109 (1998).

\bibitem{r9} H.\ Jaeger, ``The `Echo State' Approach to Analysing and
Training Recurrent Neural Networks,'' GMO Report 148, German
National Research Center for Information Technology (2001).

\bibitem{r10} W.\ Maass, T.\ Natschlager, and H.\ Markham, ``Real-Time
Computing without Stable States:  A New Framework for Neural
Computation Based on Perturbations,'' Neural Computation {\bf 14},
2531-2560 (2002).

\bibitem{r11} M.\ Luko\v{s}evi\v{c}ius and H.\ Jaeger, ``Reservoir
Computer Approaches to Recurrent Neural Network Training,''
Computer Science Review {\bf 3}, 127-149 (2009).


\bibitem{referee1} P. Antonik, M. Gulina, J. Pauwels, and S. Massar, ``Using a reservoir computer to learn chaotic attractors, with applications to chaos synchronization and cryptography,'' Phys. Rev. E 98, 012215 (2018).

\bibitem{referee2} P. Antonik, M. Haelterman, and S. Massar, ``Brain-Inspired Photonic Signal Processor for Generating Periodic Patterns and Emulating Chaotic Systems,'' Phys. Rev. Applied 7, 054014 (2017).


\bibitem{r15} N.\ F.\ Rulkov, M.\ M.\ Sushchik, L.\ S.\ Tsimring, and
H.D.T.\ Abarbanel, ``Generalized Synchronization of  Chaos in
Directionally Coupled Chaotic Systems,'' Phys.\ Rev.\ E {\bf 51},
980-994 (1995).

\bibitem{r16} L.\ Kocarev and U.\ Parlitz, ``Generalized Synchronization,
Predictability, and Equivalence of Unidirectionally Coupled
Dynamical Systems,'' Phys.\ Rev.\ Lett.\ {\bf 76}, 1816-1819
(1996).

\bibitem{r17} B.\ R.\ Hunt, E.\ Ott, and J.\ A.\ Yorke, ``Differentiable
Generalized Synchronization of Chaos,'' Phys.\ Rev.\ E {\bf 55},
4029-4034 (1997).

\bibitem{r18} A.\ E.\ Hoerl and R.\ W.\ Kennard, ``Ridge Regression:
Biased Estimation for Nonorthogonal Problems,'' Technometrics {\bf
12}, 55-67 (1970).

\bibitem{r19} R.\ Penrose, ``A Generalized Inverse for Matrices,'' Proc.\
Cambridge Philosophical Soc.\ {\bf 51}, 406-413 (1955).

\bibitem{r20} L.\  Appeltant, M.\ C.\ Soriano, G.\ van der Sande, S.\
Massar, J.\ Dambre, B.\ Schrauwen, C.\ R.\ Mirasso, and I.\
Fischer, ``Information Processing Using a Single Dynamical Node as
a Complex System,'' Nature Communications {\bf 2}, 468-473 (2013).


\bibitem{r23} L.\ Gordon and J.-P.\ Ortega, ``Reservoir Computing
Universality with Stochastic Inputs,'' IEEE Trans. on Neural
Networks and Learning Systems {\bf 23} (2019).

\bibitem{r24} E.\ N.\ Lorenz, ``Deterministic Nonperiodic Flow,'' J.\
Atmos.\ Sci.\ {\bf 20}, 130 (1963).

\bibitem{r25} H. D. Nguyen, and G. J. McLachlan, ``Maximum likelihood estimation of Gaussian mixture models without matrix operations.'', Adv. Data. Anal. Classif.  {\bf 9}:371–394 (2015).


\bibitem{r26} J.\ L.\ Kaplan and J.\ A.\ Yorke, ``Chaotic Behavior of
Multidimensional Difference Equations,'' in {\it Functional
Differential Equations and Approximations of Fixed Points}, pp.
204-227 (Springer, Heidelberg, 1979).

\bibitem{r27} J.\ D.\ Farmer, E.\ Ott, and J.\ A.\ Yorke, ``The Dimension
of Chaotic Attractors,'' Physica D {\bf 7}, 153-180 (1983).




\end{enumerate}
\end{document}